\documentstyle[aps,prb,multicol,psfig]{revtex}   
\begin{document}  
\title{Anomalous Periodicity of the Current-Phase Relationship of  
  Grain-Boundary Josephson Junctions in High-T$_c$ Superconductors}  
  
\author{E. Il'ichev, V. Zakosarenko, R.P.J. IJsselsteijn, H. E.  
  Hoenig, V. Schultze, H.-G. Meyer}  
  
\address{Department of Cryoelectronics, Institute for Physical High  
  Technology, P.O. Box 100239, D-07702 Jena, Germany}  
  
\author{M. Grajcar and R. Hlubina}  
  
\address{Department of Solid State Physics, Comenius University,  
  Mlynsk\'{a} Dolina F2, 842 15 Bratislava, Slovakia}  
  
\date{\today }  
\maketitle  
  
\begin{abstract}  
  The current-phase relation (CPR) for asymmetric 45$^\circ$ Josephson
  junctions between two $d$-wave superconductors has been predicted to
  exhibit an anomalous periodicity. We have used the single-junction
  interferometer to investigate the CPR for this kind of junctions in
  YBa$_2$Cu$_3$O$_{7-x}$ thin films.  Half-fluxon periodicity has been
  experimentally found, providing a novel source of evidence for the 
  $d$-wave symmetry of the pairing state of the cuprates. 
\end{abstract}  


\begin{multicols}{2}  
    
  There is growing evidence in favor of the $d_{x^2-y^2}$-wave
  symmetry of the pairing state of the high-temperature
  superconductors.\cite{Leggett} An unconventional pairing state
  requires the existence of zeros of the order parameter in certain
  directions in momentum space.  Thermodynamic and spectroscopic
  measurements do indeed suggest their existence, but by themselves
  they do not exclude conventional $s$-wave pairing with
  nodes.\cite{Leggett} Direct evidence for the $d$-wave pairing state
  is provided by phase-sensitive experiments, which are based on the
  Josephson effect.\cite{Tsuei} Quite generally, the current-phase
  relationship (CPR) of a Josephson junction, $I(\varphi)$, is an odd
  periodic function of $\varphi$ with a period $2\pi$. \cite{Barone}
  Therefore $I(\varphi)$ can be expanded in a Fourier series
\begin{equation}  
  I(\varphi)=I_1\sin \varphi + I_2\sin 2\varphi+\ldots .  
\label{eq:CPR}  
\end{equation}  
In the tunnel limit we can restrict ourselves to the first two terms in
Eq.~(\ref{eq:CPR}). Since the order parameter is bound to the crystal
lattice, $I(\varphi)$ of a weak link depends on the orientation of the
$d$-wave electrodes with respect to their boundary.  The existing
phase-sensitive experiments exploit possible sign changes of $I_1$ between
different geometries.\cite{Tsuei} In this Letter we present a new
phase-sensitive experimental test of the symmetry of the pairing state of
the cuprates. Namely, in certain geometries, the $I_1$ term should vanish by
symmetry. In such cases, the CPR should exhibit an anomalous periodicity.
  
Let us analyze the angular dependence of $I_{1,2}$ in a junction between two
macroscopically tetragonal $d$-wave superconductors.  As emphasized in
Ref.~\onlinecite{Walker}, also heavily twinned orthorhombic materials such
as YBa$_2$Cu$_3$O$_{7-x}$ belong to this class, if the twin boundaries have
odd symmetry.  We consider first an ideally flat interface between the
superconducting electrodes.  Let $\theta_1$ ($\theta_2$) denote the angle
between the normal to the grain boundary and the $a$ axis in the electrode 1
(2), see Fig.~1. If we keep only the lowest-order angular harmonics, the
symmetry of the problem dictates that \cite{Walker}
\begin{equation}  
I_1=I_c\cos 2\theta_1\cos 2\theta_2+  
I_s\sin 2\theta_1\sin 2\theta_2.  
\label{eq:angles}  
\end{equation}  
The coefficients $I_c,I_s$ are functions of the barrier strength,
temperature $T$, etc.  The $I_{2}$ term results from higher-order
tunneling processes and we neglect its weak angular dependence.  It is
seen from Eq.~(\ref{eq:angles}) that the criterion for the observation
of an anomalous CPR, $I_1=0$, is realized for an asymmetric 45$^\circ$
junction, i.e. a junction with $\theta_1=45^\circ$ and $\theta_2=0$.
For an interface which is not ideally flat, $\theta_i=\theta_i(x)$ are
functions of the coordinate $x$ along the junction. $I_1=0$ remains
valid even in this case, if the average values
$\langle\theta_1(x)\rangle = 45^\circ$ and $\langle\theta_2(x)\rangle
= 0$.
\begin{figure} 
\begin{center} 
\begin{tabular}{c} 
\psfig{file=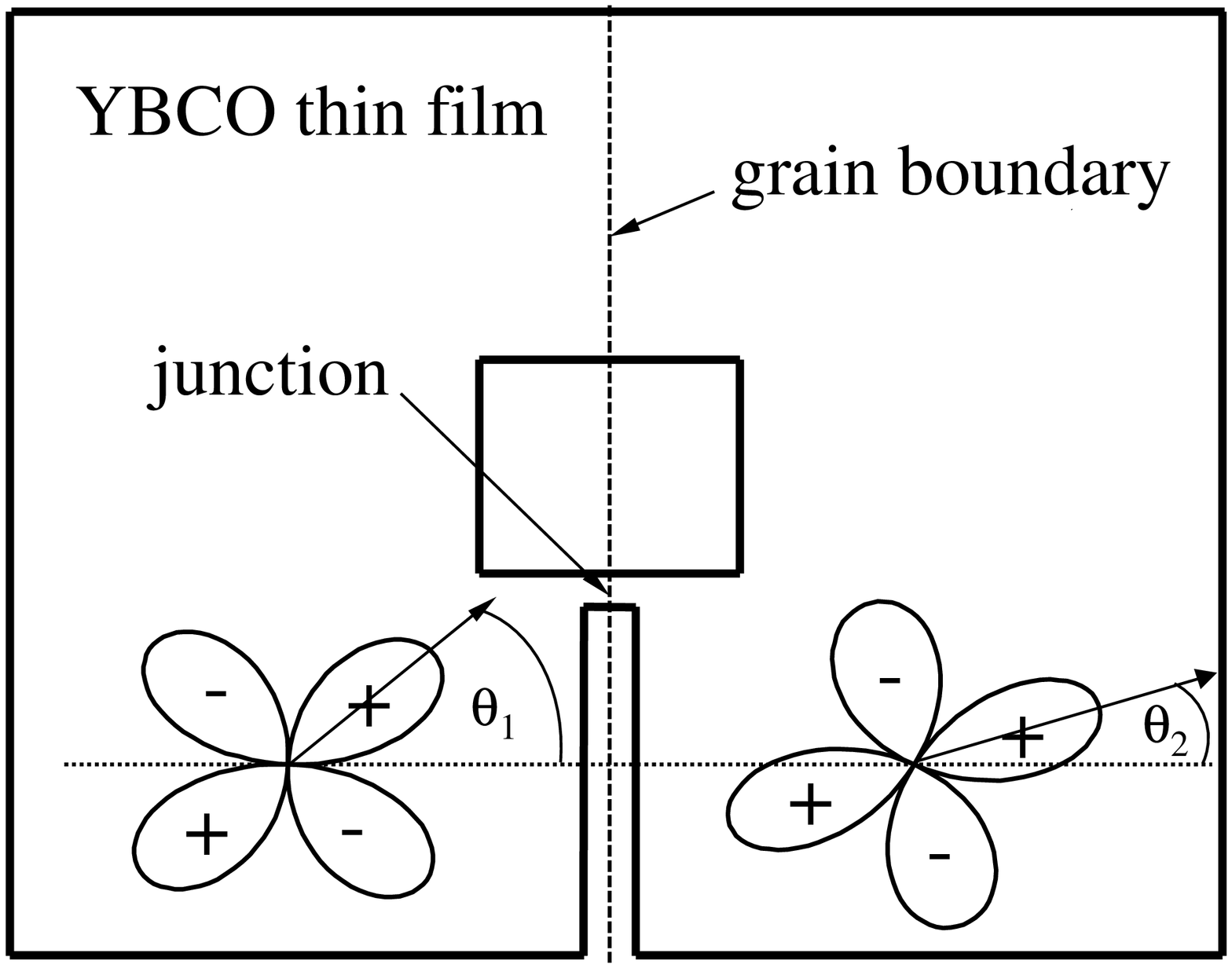,width=6cm,angle=-90} 
\end{tabular} 
\end{center} 
 \noindent   
{\small FIG.~1. Sketch of the interferometer (not in scale).}  
\end{figure}  
The $I_{2}$ term is present also in weak links based on conventional  
$s$-wave superconductors but for all known types of weak links  
$|I_2/I_1|<1$.  For instance, for a tunnel junction $|I_2/I_1|\ll 1$.  
For a SNS junction, $I\propto \sin \varphi/2$ at $T=0$,\cite{Kulik} and the
Fourier expansion Eq.~(\ref{eq:CPR}) leads to $I_2/I_1=-2/5$.  Therefore, a
possible experimental observation of $|I_2/I_1|\gg 1$ in an asymmetric
45$^\circ$ junction provides direct evidence of $d$-wave symmetry of the
pairing state in the cuprates.
  
We have investigated the CPR of YBa$_2$Cu$_3$O$_{7-x}$ thin film
bicrystals with asymmetric 45$^\circ$ [001]-tilt grain boundaries
\cite{Kirtley,Hilgenkamp,Mannhart} as sketched in Fig.~1, using a
single-junction interferometer configuration in which the Josephson
junction is inserted into a superconducting loop with a small
inductance $L$. In a stationary state without fluctuations, the phase
difference $\varphi$ across the junction is controlled by applying
external magnetic flux $\Phi_e$ penetrating the loop: $\varphi =
\varphi_e - \beta f(\varphi)$.  Here $\varphi_e=2\pi\Phi_e/\Phi_0$ is
the external flux normalized to the flux quantum $\Phi_0 = 2.07\times
10^{-15}$ Tm$^2$.  The CPR is written as $I(\varphi)=I_0 f(\varphi)$,
where $I_0$ is the maximal Josephson current.  $\beta=2\pi
LI_0/\Phi_0$ is the normalized critical current.  In order to obtain
the CPR for the complete phase range $-\pi\le\varphi\le\pi$ the
condition $\beta<1$ has to be fulfilled, because for $\beta >1$ the
curve $\varphi(\varphi_e)$ becomes multivalued and there are jumps of
$\varphi$ and a hysteresis for a sweep of $\varphi_e$.  Following
Ref.~\onlinecite{Barone}, we express the effective inductance of the
interferometer using the derivative $f'$ with respect to $\varphi$ as
$L_{int}=L[1+1/f'(\varphi)]$.  The inductance can be probed by
coupling the interferometer to a tank circuit with inductance $L_T$,
quality factor $Q$, and resonance frequency $\omega_0$.
\cite{Ilichev} External flux in the interferometer is produced by a
current $I_{dc}+I_{rf}$ in the tank coil and can be expressed as
$\varphi_e=2\pi (I_{dc}+I_{rf})M/\Phi_0 = \varphi_{dc}+\varphi_{rf}$,
where $M^2 = k^2LL_T$, and $k$ is a coupling coefficient.  Taking into
account the quasiparticle current in the presence of voltage $V$
across the junction the phase difference is given by the relation
$\varphi = \varphi_{dc}+\varphi_{rf}-\beta f(\varphi) -2\pi\tau
(\varphi )V/\Phi_0$, where $\tau (\varphi) = L/R_J(\varphi)$ and
$R_J(\varphi)$ is the resistance of the junction . In the small-signal
limit $\varphi_{rf} \ll 1$ and in the adiabatic case $\omega \tau \ll
1$, keeping only the first-order terms, the effective inductance
$L_{eff}$ of the tank curcuit-interferometer system reads
$$ 
L_{eff}=L_T\left( 1-k^2\frac{L}{L_{int}}\right)= 
L_T\left(1-\frac{k^2\beta f'(\varphi )}{1+\beta f'(\varphi)}\right). 
$$ 
Thus the phase angle $\alpha$ between the driving current and the 
tank voltage $U$ at the resonant frequency of the tank circuit 
$\omega_0$ is 
\begin{equation}  
\label{alpha}  
\tan\alpha(\varphi)=\frac{k^2Q\beta f'(\varphi)}{1+\beta   
f'(\varphi)}.  
\end{equation}  
Using the relation $[1+\beta f'(\varphi)]d\varphi =d\varphi_{dc}$  
valid for $\varphi_{rf} \ll 1$ and $\omega\tau \ll 1$, one  can  
find the CPR from Eq.~(\ref{alpha}) by numerical integration.  
  
The advantage of the measurement of the CPR of an asymmetric 45$^\circ$
junction with respect to the by-now standard phase-sensitive tests of
pairing symmetry based on the angular dependence of $I_1$ is twofold. 
First, it avoids the complications of the analysis of experiments caused by
the presence of the term $I_s$.\cite{Walker} Second, a flux trapped in the
SQUID does not invalidate the conclusions about the ratio $|I_2/I_1|$ and
hence about the pairing symmetry, while this is not the case in standard
phase-sensitive tests of the $d$-wave symmetry of the pairing
state.\cite{Klemm}
  
The films of thickness 100 nm were fabricated using standard pulsed laser
deposition on (001) oriented SrTiO$_3$ bicrystalline substrates with
asymmetric [001] tilt misorientation angles $45^\circ\pm1^\circ$.  They were
subsequently patterned by Ar ion-beam etching into $4\times 4$ mm$^2$ square
washer single-junction interferometer structures (Fig. 1).  The widths of
the junctions were $1\div 2$ $\mu$m. The washer square holes had a
side-length of 50 $\mu$m. This geometry of the interferometer gives
$L\approx 80$ pH.  The resistance of the junction is higher than 1 $\Omega$
and the condition for the adiabatic limit $\omega \tau \ll 1$ is satisfied. 
For measurements of $\alpha(\varphi_{dc})$, several tank circuits with
inductances $0.2\div 0.8$ $\mu$H and resonance frequencies $16\div 35$ MHz
have been used.  The unloaded quality factor of the tank circuits
$70<Q<150$ has been measured at various temperatures. The coupling 
factor $k$ was determined from the period $\Delta I_{dc}$ of 
$\alpha(I_{dc})$ using $M\Delta I_{dc}=\Phi_0$. Its value varied 
between 0.03 and 0.09. The amplitude of $I_{rf}$ was set to produce 
the flux in the interferometer lower than $0.1 \Phi_0$. 
  
The measurements have been performed in a gas-flow cryostat with a
five-layer magnetic shielding in the temperature range $4.2\le T<90$ K.  The
experimental setup was calibrated by measuring interferometers of the same
size with 24$^\circ$ and 36$^\circ$ grain boundaries.  We have studied 5
samples, out of which sample No.~1 exhibited the most anomalous behavior.
Samples Nos.~2,3 were less anomalous and the remaining two samples had high
critical currents and their
$I(\varphi)$ was conventional. In Fig.~2 we plot the phase angle 
$\alpha$ as a function of the dc current $I_{dc}$ for samples 
Nos.~1,2.  The data for the 36$^\circ$ junction is shown for 
comparison.  Note that at $T=40$~K the periodicity of  
$\alpha(I_{dc})$ 
is the same for all samples. 
\begin{figure}  
\begin{center} 
\begin{tabular}{c} 
\psfig{file=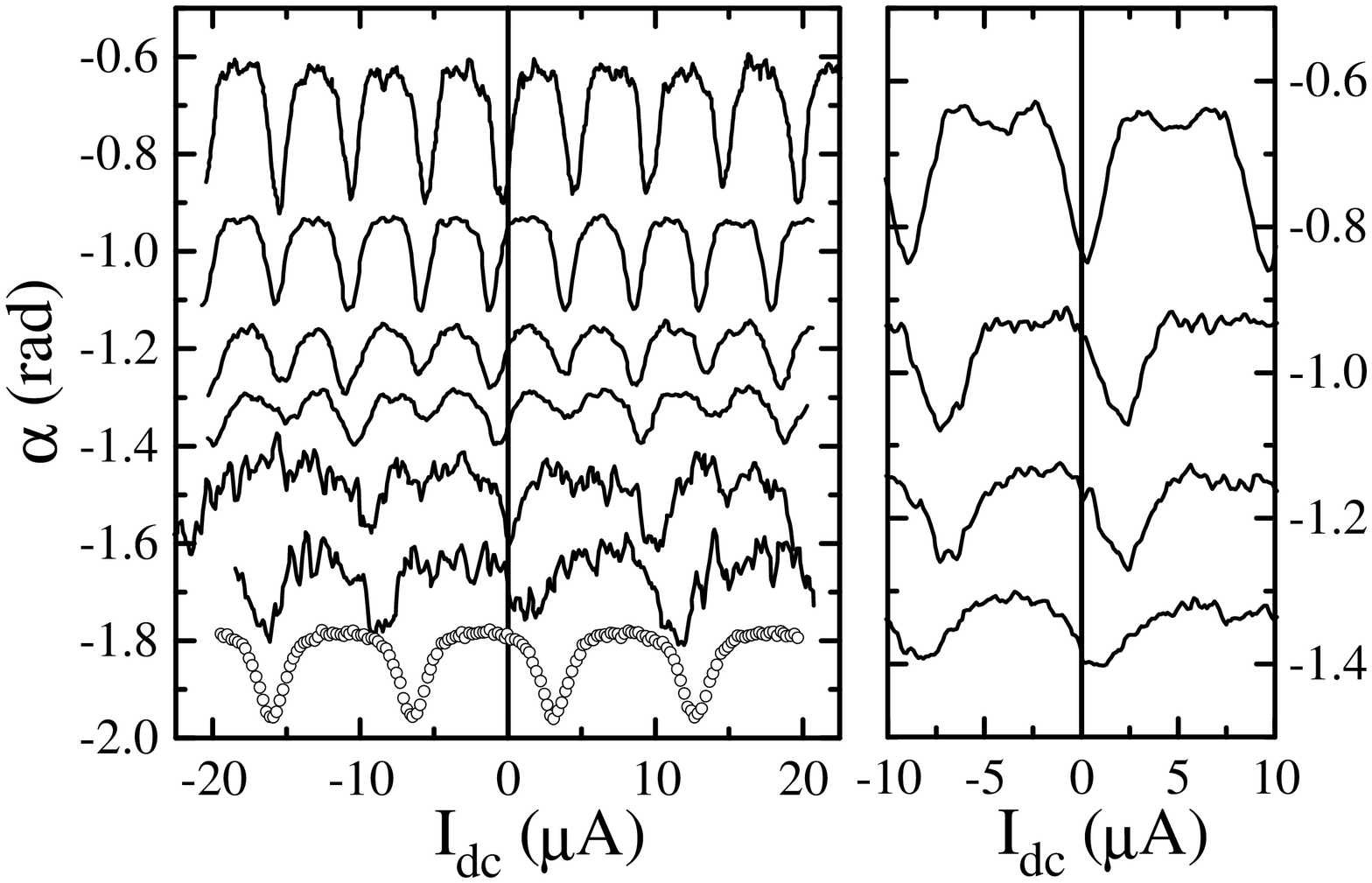,width=7cm,angle=-90} 
\end{tabular} 
\end{center}  
\noindent {\small FIG.~2. Left panel: Phase angle between the driving  
  current and the output voltage measured for the sample No. 1 at  
  different temperatures as a function of the dc current $I_{dc}$.  
  The curves are shifted along the $y$ axis and the data for $T=$ 30  
  and 40 K are multiplied by factor 4 for clarity. From top to  
  bottom, the data correspond to $T=$4.2, 10, 15, 20, 30 and 40 K.
  The data measured on 36$^\circ$ bicrystals ($\theta_1\approx 36^\circ ,  
  \theta_2\approx 0$) at $T=$40 K in the same washer geometry are  shown  
  for comparison (open circles). Right panel: The same for the sample No.~2.
  From top to bottom, the data correspond to $T=$4.2, 10, 15 and 20 K.}  
\end{figure}  
We assume that the period of $\alpha(I_{dc})$ at $T=$40 K, $\Delta
I_{dc}=$9.6 $\mu$A, corresponds to $\Delta\varphi_{dc}=2\pi$.  In
order to determine the CPR we take $\varphi_{dc}=0$ at a maximum or
minimum of $\alpha$. This is necessary in order to satisfy
$I(\varphi=0)=0$, as required by general principles.\cite{Barone} The
experimentally observed shift of the first local extreme of
$\alpha(I_{dc})$ from $I_{dc}=0$ (Fig. 2) can be due to flux trapped
in the interferometer washer. Most probably, this flux resides in the
long junction of the interferometer. The long junction does not play
an active role because the Josephson penetration depth is much shorter
than its length, and external fields producing $I_{dc}$ are smaller
than its critical field.  Nevertheless the long junction sets the
phase difference for $I_{dc}=0$ at the small junction.
\begin{figure}  
\begin{center} 
\begin{tabular}{c} 
  \psfig{file=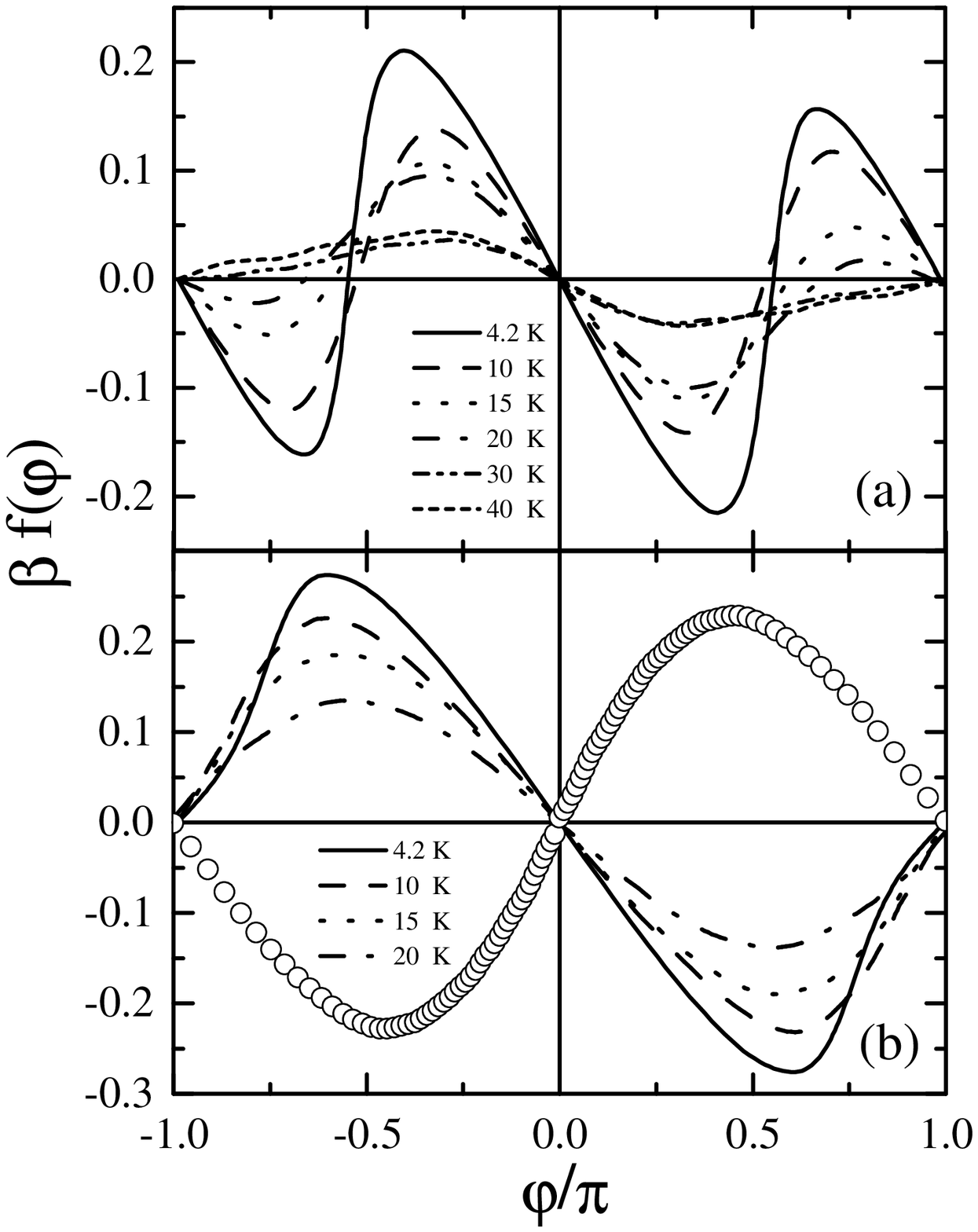,width=7cm}  
\end{tabular} 
\end{center} 
\noindent  
{\small FIG.~3. a) Josephson current through the junction for the
  sample No. 1 as a function of the phase difference $\varphi$,
  determined from the data in Fig.~2. The statistics of
  $\alpha(\varphi)$ was improved by folding the data back to the
  interval $\langle 0,\pi\rangle$ and taking an average. The symmetry
  $\alpha(\varphi)=\alpha(-\varphi)$ was assumed. b) The same for the
  sample No.~2. The data for the asymmetric 36$^\circ$ bicrystal at
  $T=40 $ K (open circles) is also shown.}
\end{figure}  
In Fig.~3, we show the CPR determined from the data in Fig.~2.  For
all curves we have performed a minimal necessary shift consistent with
$I(\varphi=0)=0$. Thus, for the samples Nos.~1,2 we have assumed that
at $\varphi_{dc}=0$ a minimum of $\alpha(\varphi_{dc})$ is realized.
For an interferometer with a conventional $s$-wave weak link (and also
for the 36$^\circ$ junction), at $\varphi_{dc}=0$ a maximum of
$\alpha(\varphi_{dc})$ is realized.  Note that the maximum (minimum)
of $\alpha(\varphi_{dc})$ at $\varphi_{dc}=0$ implies a diamagnetic
(paramagnetic) response of the interferometer in the limit of small
applied fields.  In Fig.~4 we plot the coefficients $I_1$ and $I_2$
determined by Fourier analysis of the CPR for the sample No.~1 at
various temperatures. With decreasing $T$, $|I_2|$ grows monotonically
down to $T=4.2$ K, while the $I_1$ component exhibits only a weak
temperature dependence.
  
Our experimental results can be understood as follows.  Deviations
from ideal geometry of the asymmetric 45$^\circ$ junction,
$\langle\theta_1\rangle=45^\circ+\alpha_1$ and
$\langle\theta_2\rangle=\alpha_2$, lead to a finite value of $I_1$.
Thus, imperfections of the junction increase its critical current. For
this reason we believe that samples Nos.~2-5 contain imperfections and
from now we concentrate on nearly ideal junctions (such as sample
No.~1) with $|\alpha_1|,|\alpha_2|\ll 1$.  For such junctions, the
ratio $I_2/I_1$ exhibits the following temperature dependence.  For
$T\rightarrow 0$, $|I_2/I_1|\gg 1$.  The region $T\sim T_c$ can be
analyzed quite generally within Ginzburg-Landau theory.  Let the
electrodes be described by (macroscopic) order parameters
$\Delta_{1,2}=|\Delta| e^{i\varphi_{1,2}}$.  Then the phase-dependent
part of the energy of the junction is $E=a[\Delta_1\Delta_2^\star+{\rm
  H.C.}]  +b[(\Delta_1\Delta_2^\star)^2+{\rm H.C.}]+\ldots$ where
$a,b,\ldots$ depend weakly on $T$.\cite{Huck} Thus for $T$ close to
$T_c$ we estimate $I_1\propto |\Delta|^2\propto (T_c-T)$ and
$I_2\propto |\Delta|^4\propto (T_c-T)^2$, leading to $|I_2/I_1|\ll 1$.
These expectations are qualitatively consistent with the experimental  
data shown in Fig.~4.  
\begin{figure}  
\begin{center} 
\begin{tabular}{c} 
\psfig{file=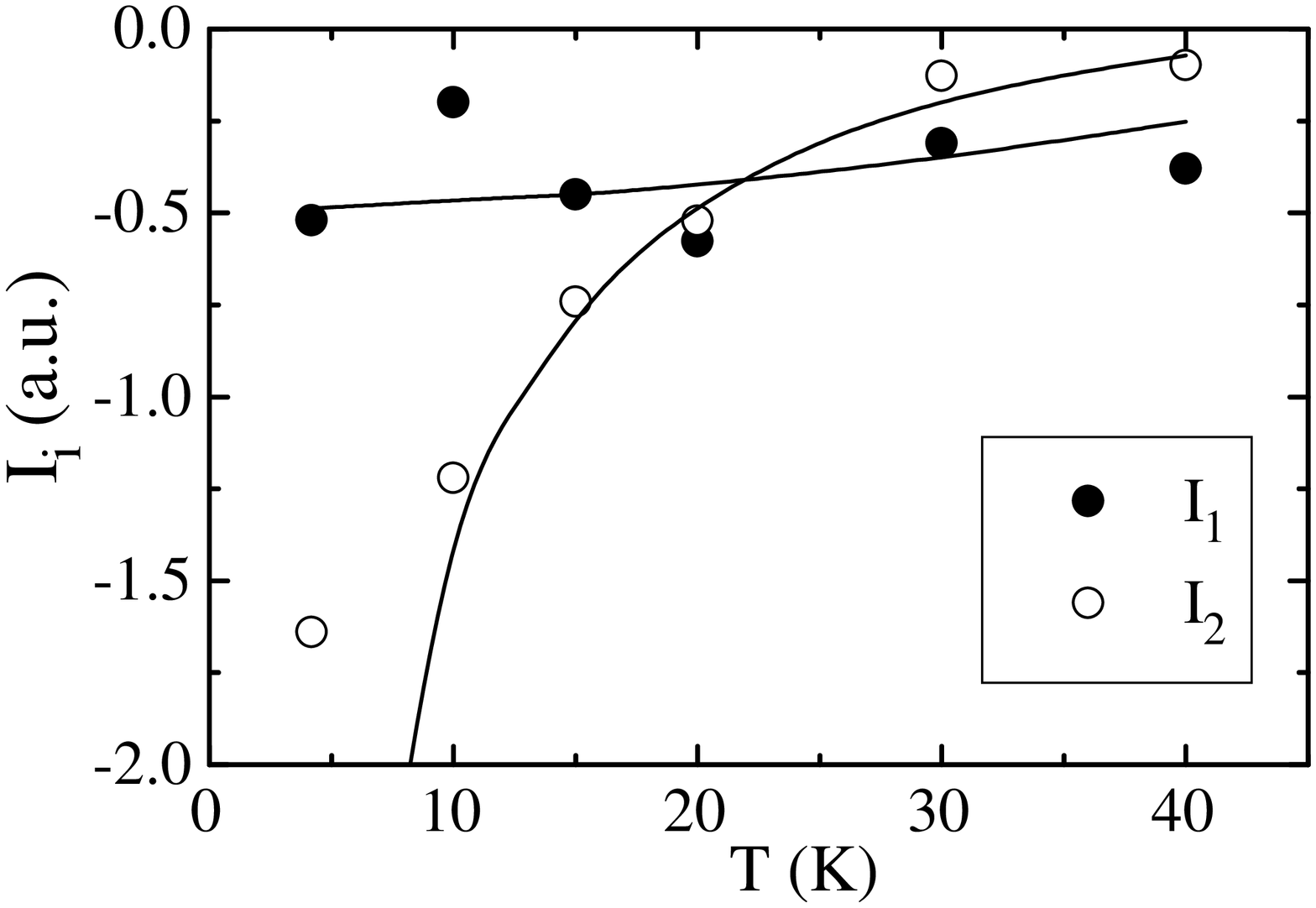,width=7cm,angle=-90}  
\end{tabular} 
\end{center} 
 \noindent  
 \tighten {\small FIG.~4.  Temperature dependence of the Fourier  
   expansion coefficients $I_{1,2}$ determined from the experimental  
   data in Fig.~3a. Solid lines are the Fourier expansion coefficients  
   for the numerical data in Fig.~5.}  
\end{figure}  
So far, our discussion was based solely on symmetry arguments. Let us
attempt a more quantitative analysis of our data now.  Two different
microscopic pictures of asymmetric 45$^\circ$ Josephson junctions
between $d$-wave superconductors have been considered in the
literature. The first picture assumes a microscopically tetragonal
material and an ideally flat interface.\cite{Huck,Tanaka,Barash}
Within this picture, there are two contributions to the Josephson
current.  The first is due to bulk states and in the tunnel limit it
is well described by the Sigrist-Rice term $I_c$ in
Eq.~(\ref{eq:angles}).\cite{Sigrist} The second is due to mid-gap
states which develop close to the surfaces of unconventional
superconductors.\cite{Hu} $I(\varphi)$ for the sample No.~1 calculated
according to the model of Ref.~\onlinecite{Tanaka} is shown in Fig.~5.
The experimental data can be fitted by a relatively broad range of
barrier heights.  However, if we require the 36$^\circ$ junction to be
fitted by the same (or smaller) barrier height as for the 45$^\circ$
junction, we conclude the barrier must be rather low.\cite{note} The
$T$ dependence of $I(\varphi)$ requires a choice of $T_c\approx 60$ K
in the non-selfconsistent theory of Ref.~\onlinecite{Tanaka}.  The
reduction from the bulk $T_c=90$ K is probably due to a combined
effect of surface degradation and order-parameter suppression at the
sample surface.  The temperature dependence of the ratio of the $\pi$
and $2\pi$ periodic components in $I(\varphi)$ is seen to be in
qualitative agreement with experimental data in Fig.~3a.  This is
explicitly demonstrated in Fig.~4 where we compare the experimentally
obtained $I_{1,2}$ with the results of the Fourier analysis of the
curves in Fig.~5. The divergence of $I_2$ as $T\rightarrow 0$ is an
artifact of the ideal junction geometry assumed in
Ref.~\onlinecite{Tanaka}. If the finite roughness of the interface is
taken into account, this divergence is cut off and the experimental
data in Fig.~4 do indeed resemble theoretical predictions for a rough
interface.\cite{Barash} However, the nonselfconsistent theory of
Ref.~\onlinecite{Tanaka} is unable to explain the experimentally
observed steep CPR close to the minima of the junction energy (see
Fig. 3a). In the limit of vanishing barrier height, the theoretical
CPR does have steep portions, but these are located close to the
maxima of the junction energy (see also Ref.~\onlinecite{Huck}).
\begin{figure}  
\begin{center} 
\begin{tabular}{c} 
\psfig{file=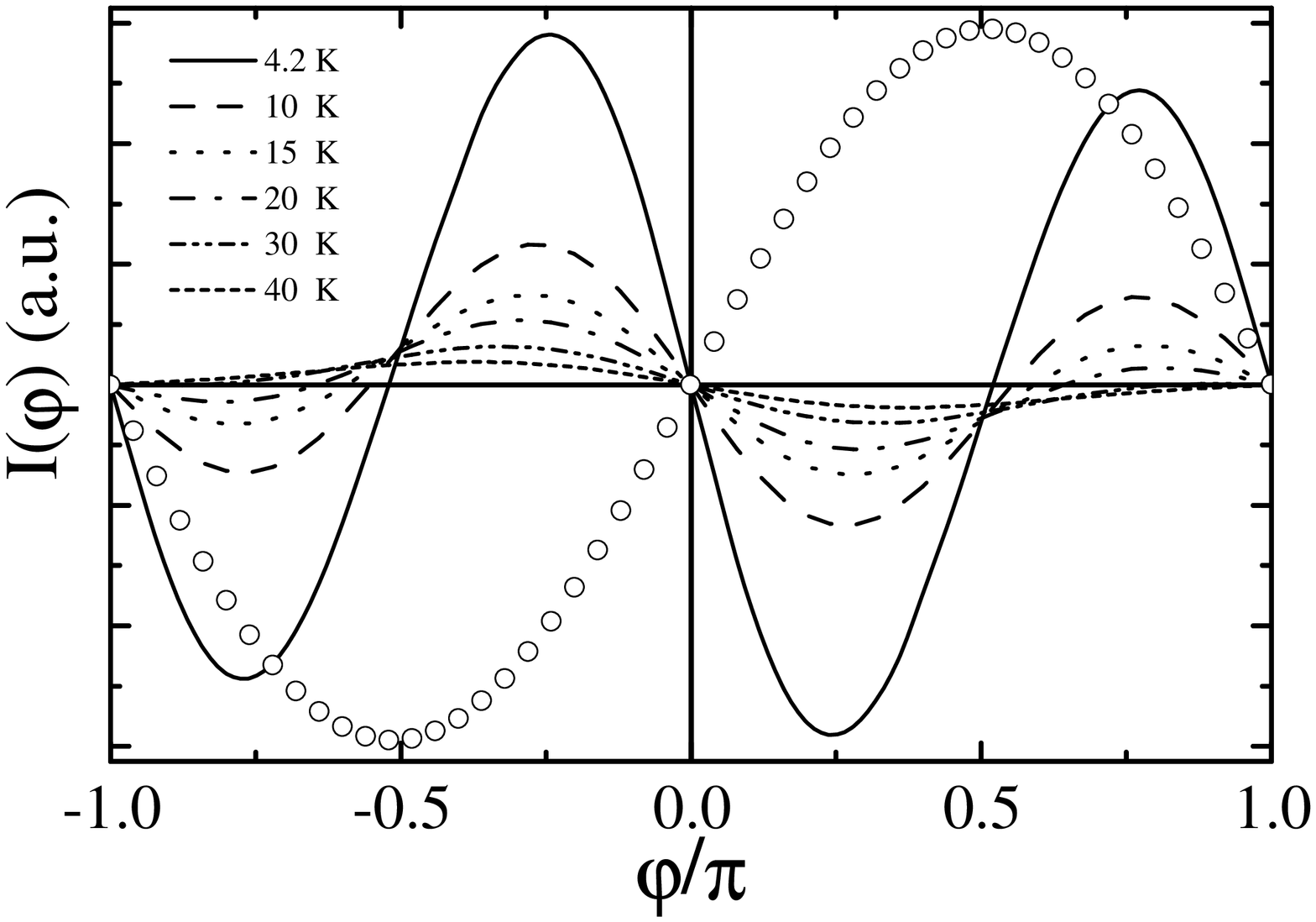,width=7cm,angle=-90}   
\end{tabular} 
\end{center} 
\noindent   
{\small FIG. 5.  $I(\varphi)$ calculated according to Eq. (64) of  
  Ref.~\onlinecite{Tanaka} for a junction with $\theta_1=45.5^\circ$,  
  $\theta_2=0$, $\lambda d=1.5$, $\kappa=0.5$, and $T_c=60$ K.  
  $I(\varphi)$ at $T=40$ K for the 36$^\circ$ bicrystal (open circles)  
  was calculated for the same parameters except for $\theta_1=36^\circ$.}  
\end{figure}  
  
In a different approach to the asymmetric 45$^\circ$ junction, one
assumes a heavily twinned orthorhombic material (which is
macroscopically tetragonal, however) and/or a meandering interface
with $\theta_i=\theta_i(x)$.\cite{Millis,Mints} Hence the critical
current density $j_c(x)$ is a random function with a typical amplitude
$\langle|j_c(x)|\rangle\sim j_c$. If the average critical current
along the junction $\langle j_c\rangle\ll j_c$, a spontaneous flux is
generated in the junction, and $|I_2/I_1|\gg 1$.\cite{Millis,Mints} In
particular, for $\langle\theta_1\rangle=45^\circ$ and
$\langle\theta_2\rangle=0$, there is an equal amount of parts having
positive and negative $j_c$, leading to $\langle j_c\rangle=0$ and
$I_1=0$.  Note that also within the picture of
Refs.~\onlinecite{Millis,Mints}, the $d$-wave symmetry of the pairing
state is crucial, otherwise the condition $\langle j_c\rangle\ll j_c$
is difficult to satisfy.
  
Our present understanding of $I(\varphi)$ in the asymmetric 45$^\circ$
junction is only qualitative.  There is considerable experimental
evidence \cite{Kirtley,Hilgenkamp,Mannhart} that the grain boundary
junctions are at most piecewise flat. However, we cannot say whether
the shape of $I(\varphi)$ is dominated by the mid-gap states in the
microscopically flat regions, or by spontaneous flux generation due to
the spatial inhomogeneity of the junction. This issue requires further
study.
  
In conclusion, we have measured the magnetic field response of a
single-junction interferometer based on asymmetric 45$^\circ$
grain-boundary junctions in YBa$_2$Cu$_3$O$_{7-x}$ thin films.
Half-fluxon periodicity has been experimentally found, in agreement
with theoretical predictions for $d_{x^2-y^2}$-wave superconductors.
Hence, our results provide a novel source of evidence for the $d$-wave
symmetry of the pairing state in the cuprates.
  
Financial support by the DFG (Ho 461/1-1) is gratefully acknowledged.
M. G. and R. H. were supported by the Slovak Grant Agency Grant No.
1/4300/97 and the Comenius University Grant No.  UK/3927/98.

\end{multicols}


\begin{references}
\bibitem{note}This is consistent with the Fourier analysis of the data
  in Fig.~3 which finds non-negligible $I_n$ also for $n\geq 3$.
\end{references}
\end{document}